\documentclass[12pt]{article}
\usepackage{graphicx}

\begin{document}

\def\onehalf{{\textstyle \frac12}}
\def\ii{{\rm i}}
\def\dd{{\rm d}}
\def\jour#1#2#3#4{{\it #1{}} {\bf #2}, #3 (#4)}
\def\lab#1{\label{eq:#1}}
\def\rf#1{(\ref{eq:#1})}
\def\Lie#1{\hbox{\sf #1}}
\def\rectangulo#1{\centerline{\framebox{\LARGE #1}}}
\def\ket#1{\,\vert{#1}\rangle}
\def\bra#1{\langle{#1}\vert}
\def\tsty#1#2{{\textstyle\frac{#1}{#2}}}
\def\ssty#1{{{\scriptscriptstyle #1}}}

\newcommand{\be}{\begin{equation}}
\newcommand{\ee}{\end{equation}}
\newcommand{\bea}{\begin{eqnarray}}
\newcommand{\eea}{\end{eqnarray}}

\begin{center}
{\LARGE Discrete Bessel and Mathieu functions} \\[20pt] 	
 {Kenan Uriostegui}\footnote{Posgrado en Ciencias F\'{\i}sicas, 
Universidad Nacional Aut\'onoma de M\'exico} and {Kurt Bernardo Wolf}\\[15pt]
{Instituto de Ciencias F\'{\i}sicas\\
Universidad Nacional Aut\'onoma de M\'exico\\Av.\ Universidad s/n, 
Cuernavaca, Morelos 62251, M\'exico}
\end{center} 

\vskip25pt

\begin{abstract}
The two-dimensional Helmholtz equation separates in
elliptic coordinates based on two distinct foci, a limit case 
of which includes polar coordinate systems when the two foci 
coalesce. This equation is invariant under the Euclidean
group of translations and orthogonal transformations; we replace 
the latter by the discrete dihedral group of $N$ discrete rotations 
and reflections. The separation of variables in polar and elliptic 
coordinates is then used to define {\it discrete\/} Bessel and Mathieu 
functions, as approximants to the well-known {\it continuous\/} 
Bessel and Mathieu functions, as $N$-point Fourier
transforms approximate the Fourier transform over the circle,
with integrals replaced by finite sums. We find that these
`discrete' functions  approximate the numerical values of 
their continuous counterparts very closely and preserve 
some key special function relations.
\end{abstract}


\section{Introduction}             \label{sec:one}  

The role of the Euclidean group of translations, reflections and rotations 
in the determination of the coordinate systems that separate the solutions
of the two-dimensional Helmholtz equation is well known from the 
work by Willard Miller Jr.\ \cite[Ch.\ 1]{Miller-book}. This symmetry accounts 
for their separability in four coordinate systems: Cartesian, polar, parabolic 
and elliptic. Only the elliptic system is generic; when the two foci coalesce,
this system becomes the polar one with angular and radial coordinates; 
when one focus departs to infinity the system becomes parabolic;
and when both foci do, it becomes Cartesian. 

The polar decomposition was used by Biagetti et al.\ \cite{Biagetti-etal}
to first introduce a discrete version of Bessel functions based on an 
expansion of plane waves into a finite number of polar components 
---that was not quite complete. This was properly completed in Ref.\ \cite{UW1}, 
defining {\it discrete Bessel functions\/} $B_n^\ssty{N}(\rho)$, which 
approximate the usual {\it continuous\/} Bessel functions 
$J_n(\rho)$ by replacing Fourier series over a circle ${\cal S}^1$ by 
the finite Fourier transform on $N$ equidistant points on that circle,
\be 
	\theta_m = 2\pi m/N, 
		\qquad  m\in\{0,1,\ldots,N{-}1\}=: {\cal S}^1_\ssty{(N)},
			\lab{k-l-Z}
\ee 
where $m$ is counted modulo $N$.
It was found that these discrete functions approximated very closely
(of the order $10^{-16}$) the corresponding continuous ones over a 
region, roughly $0\le n+\rho < N$. 

Several authors have introduced functions that approximate 
the well-known continuous Bessel functions $J_n(\rho)$, for the purpose 
of reducing computation time, or to provide new classes
of solutions to difference equations that will share some of 
their salient properties \cite{RHBoyer,Bohner-Cuchta,Slavik}.
Our approach follows the well known approximation
afforded by the $N$-point finite Fourier transform to the
integral Fourier transform over the circle.
This is done for polar and elliptic coordinates, and introduces
both `discrete' Bessel and Mathieu functions. These functions, 
we should emphasize, differ from those proposed in the works cited 
above, which are also distinct in definition and purpose among
themselves. By construction it will follow that under $N\to\infty$,
these discrete functions become the continuous ones, although this
limit requires further mathematical precision, as it may involve 
Gibbs-type oscillation phenomena that we cannot address here.

In Sect.\ \ref{sec:two} we present this discretization method
and a resum\'e of the results in Ref.\ \cite{UW1} for Bessel 
functions, to note that the discrete functions thus defined 
approximate the continuous ones remarkably well. 
In Sect.\ \ref{sec:three} of the present paper we 
apply the strategy of replacing harmonic analysis on ${\cal S}^1$
by ${\cal S}^1_\ssty{(N)}$ to define discrete approximants to 
the Mathieu functions of first and second kind in the elliptic 
coordinate system. All relations are backed by
numerical verification. In the concluding Sect.\ \ref{sec:four} 
we provide some further connections and preliminary conclusions.


\section{Continuous and discrete Bessel functions}
							\label{sec:two}

The Helmholtz equation for wavefields $f(x,y)$ of (fixed) real
wavenumber $\kappa\in{\cal R}$, is
\be
	(\partial_x^2+\partial_y^2+\kappa^2)f(x,y)=0, \lab{Helm-eq}
\ee
with $\partial_z\equiv\partial/\partial_z$ and $(x,y)\in{\cal R}^2$. 
In this section we follow the well known case of polar coordinates, 
\be
	x=r\cos\theta,\quad  y=r\sin\theta, \qquad 
		r\in[0,\infty),\quad \theta\in(-\pi,\pi]={\cal S}^1.
	\lab{coord-rad}
\ee
A key assumption is a Hilbert space structure for 
the solutions $f(x,y)$ by which one can write them as
the two-dimensional Fourier transform, 
\be 
	f(x,y) = \frac1{2\pi}\int\!\!\!\int_{{\cal R}^2}\dd\kappa_x\,\dd\kappa_y\,
		\exp\ii(x\kappa_x+y\kappa_y)\,\widetilde{f}(\kappa_x,\kappa_y)
		\lab{Helm-Fou}.
\ee
The Helmholtz equation \rf{Helm-eq} is then correspondingly transformed to 
a conjugate space $(\kappa_x,\kappa_y)\in{\cal R}^2$ where it reads
$(\kappa^2-\kappa_x^2-\kappa_y^2)\widetilde{f}(\kappa_x,\kappa_y)=0$,
which we can also refer to polar coordinates $\kappa_x=\kappa\cos\phi$, 
$\kappa_y=\kappa\sin\phi$, with the surface element 
$\dd\kappa_x\,\dd\kappa_y=\kappa\,\dd\kappa\,\dd\phi$.
The solutions to the Fourier-transformed Helmholtz equation are thus 
reduced by a Dirac $\delta$-distributions in the radius 
\cite[Ch.\ 1]{Miller-book}, as
$\widetilde{f}(\kappa_x,\kappa_y)=\sqrt{2\pi}\kappa^{-1}
\delta(\kappa-\widetilde\kappa)\,{f}_{\!\circ}(\phi)$,
with ${f}_{\!\circ}(\phi)$ a function on the $\phi$-circle ${\cal S}^1$ of 
radius $\widetilde\kappa$, that we write again $\kappa$, understanding that it 
is the fixed wavenumber. 
The Helmholtz solutions \rf{Helm-Fou} thus acquire the single-integral form 

\be 
	f(x,y) = \frac1{\surd 2\pi}\int_{{\cal S}^1}\dd\phi\,\exp\ii\kappa(x\cos\phi+y\sin\phi)
		\,{f}_{\!\circ}(\phi), \lab{rad-Helm}
\ee
with the Hilbert space structure based on the inner product
of functions $f^\ssty{(1)}_{\!\circ}(\phi)$ and $f^\ssty{(2)}_{\!\circ}(\phi)$ on the circle, 
\be 
	(f^\ssty{(1)}_{\!\circ},f^\ssty{(2)}_{\!\circ})_\circ:=\int_{{\cal S}^1}\dd\phi\,
		f^\ssty{(1)}_{\!\circ}(\phi)^*f^\ssty{(2)}_{\!\circ}(\phi).
				\lab{inn-prod-c}
\ee

It is here that we reduce the continuous circle Fourier transform to the $N$-point 
discrete Fourier transform, from ${{\cal S}^1}$ to ${\cal S}^1_\ssty{(N)}$, 
replacing integrals by summations and the continuous variable $\phi\in{\cal S}^1$
with $\phi_m\in{\cal S}^1_\ssty{(N)}$, as
\be 
	\int_{{\cal S}^1}\dd\phi\,{F}_{\!\circ}(\phi)
	 \leftrightarrow \sum_{m\in{\cal S}^1_\ssty{(N)}}{F}(\phi_m), \qquad
		{2\pi \leftrightarrow N, \atop 	\phi_m=2\pi m/N, }   \lab{repl-lim}
\ee
for $m\in\{0,1,\ldots,N{-}1\}$ counted modulo $N$; the set of $N$ discrete 
angles $\phi_m$ are thus equidistant by $2\pi/N$.  
The functions $f(\phi_m) \equiv f_m$  can be interpreted as sample points of a 
continuous function, or as the index for the list of components
of an $N$-cyclic vector. In either case, the inner product of two 
discrete functions $f^\ssty{(1)}_n$ and $f^\ssty{(2)}_n$ is naturally
\be 
	(f^\ssty{(1)},f^\ssty{(2)})_\ssty{(N)} := \sum_{n=0}^{N-1} 
		f_n^{\ssty{(1)}*}\,f^\ssty{(2)}_n,
			\lab{disc-inn-prod}
\ee
and it is clear that the $N\to\infty$ limit will lead back from the
discrete to the continuum, with the approximations and limits familiar
from Fourier theory. 

The Helmholtz equation \rf{Helm-eq} in polar coordinates,
multiplied by $r^2$, 
\be 
	(r^2\partial^2_r + r\partial_r + \partial^2_\phi 
		+ \kappa^2)f(r,\phi)=0,
		\lab{Helm-eq-rad}
\ee
shows that solutions can be factored into a
function of the radius times a function of the angle
as $f(r,\phi)=R(r)\,\Phi(\phi)$, while \rf{rad-Helm}
implies that solutions $\Phi(\phi)$ for the angular
factor will determine a corresponding radial factor
$R(r)$. An orthonormal and complete set of 
eigenfunctions of $\partial^2_\phi$ over the circle 
$\phi\in{\cal S}^1$ is the set of phases $\Phi_n(\phi) := 
(2\pi)^{-1/2}\exp(\ii n\phi)$, with integer 
$n\in\{0,\pm1,\ldots\}$, and inner products
$(\Phi_n,\Phi_{n'})_\circ=\delta_{n,n'}$.
When the domain of these functions is restricted 
from $\phi\in{\cal S}^1$ to 
$\phi_m\in{\cal S}^1_\ssty{(N)}$ as in \rf{k-l-Z},
we retain the subset of $N$ functions 
on the $N$ points in ${\cal S}^1_\ssty{(N)}$,  
given by
\be 
	\Phi^\ssty{(N)}_n(\phi_m):=\frac1{\surd N}\exp(\ii n\phi_m)
		=\frac1{\surd N}\exp\bigg(\frac{2\pi\ii m n}N\bigg) 
		= \Phi^\ssty{(N)}_{n\pm N}(\phi_m),
					\lab{base-Phi}
\ee
labeled by the {\it cyclic\/} subset $n\in\{0,1,\ldots,N{-}1\}$, 
that are also orthonormal under the common inner product 
\rf{disc-inn-prod} for discrete functions on ${\cal S}^1_\ssty{(N)}$,
and complete:
\be
	(\Phi^\ssty{(N)}_n,\Phi^\ssty{(N)}_{n'})_\ssty{(N)}= \delta_{n,n'},\qquad
		\sum_{n=0}^{N-1} \!\Phi^\ssty{(N)}_n(\phi_m)^*\,
			\Phi^\ssty{(N)}_{n}(\phi_{m'})=\delta_{m,m'}.
				\lab{ortho-complete}
\ee
Returning to \rf{rad-Helm} with $(x,y)$ in the polar coordinates $(r,\theta)$ 
of \rf{coord-rad}, and taking for ${f}_\circ(\phi_m)$ the basis functions 
\rf{base-Phi} on the discrete points of ${\cal S}^1_\ssty{(N)}$, we write the 
$N$ solutions to the discretized Helmholtz equation, labeled 
by cyclical $n\in\{0,1,\ldots,N{-}1\}$, as
\be 
	\begin{array}{rcl}   {}\!\!\!\!\!
	f_n(r,\theta_k)\!\!\!&=&\!\!\!\displaystyle\frac1{\surd N}\sum_{m\in{\cal S}^1_\ssty{(N)}}
		\!\!\!\exp[\ii\kappa r(\cos\theta_k\cos\phi_{m}+\sin\theta_k\sin\phi_{m})]
		\,\Phi^\ssty{(N)}_n(\phi_{m})  \\
	{}\!\!\!&=&\!\!\!\displaystyle\frac1{N}\sum_{m\in{\cal S}^1_{(N)}}
			\!\!\!\exp[\ii\kappa r\cos(\theta_k-\phi_{m})] \exp(\ii n \phi_{m})\\
	{}\!\!\!&=&\!\!\!\displaystyle\frac{e^{\ii n (\theta_k+\pi/2)}}N\sum_{m\in{\cal S}^1_\ssty{(N)}}
			\!\!\!\exp(\ii\kappa r\sin\varphi_m) \exp(-\ii n \varphi_{m}),\end{array}
			\lab{disc-Helm}
\ee
having replaced $\varphi_m:=\theta_k-\phi_m+\frac12\pi$ in the summation 
over the $N$ discrete points on the circle. 

Following Miller \cite[p.\ 29]{Miller-book}, the phase in front of 
\rf{disc-Helm}, $e^{\ii n\theta_n(\theta_k+\pi/2)} = \ii^n e^{2\pi\ii nk /N} 
= \ii^n\sqrt{N}\Phi^\ssty{(N)}_n(\theta_k)$, is extracted 
to write the functions as
\be
	f_n(r,\theta_k)= \ii^n \sqrt{N}\,B^\ssty{(N)}_n(\kappa r)\, \Phi^\ssty{(N)}_n(\theta_k),
		\lab{fase-B}
\ee
where the radial factor $B^\ssty{(N)}_n(\rho)$, $\rho:=\kappa r$, are the 
{\it discrete Bessel functions}. From \rf{disc-Helm} these functions are 
seen to be {\it real\/} and their parities, using coefficients 
$\{c_n,s_n\}:=\{1,0\}$ for $n$ even or $\{0,1\}$ for $n$ odd, 
can be written as
\be  
	\begin{array}{rcl}
	B^\ssty{(N)}_n(\rho)&=&\displaystyle
		\frac1N \sum_{m\in{\cal S}^1_{N}}
		\!\!\! \exp(\ii\rho \sin\varphi_m)\,
			[c_n\cos(n\varphi_m)-\ii s_n\sin(n\varphi_m)]\\
	&=& \displaystyle \frac1N \sum_{m\in{\cal S}^1_{N}}
		\!\!\! \exp(\ii\rho \sin\varphi_m) \times
			\left\{ \begin{array}{rl}
					\cos n\varphi_m,& n\hbox{ even},\\				
					-\ii\sin n\varphi_m,& n\hbox{ odd}.
			\end{array} \right. \end{array}  \lab{Bdef-int}
\ee
The distinction between even and odd cases of $n$, 
as done in \cite{UW1}, is subtle but important to obtain 
the correct result for all $n$'s ({\it cf}.\ \cite[Eq.\ (9)]{Biagetti-etal}).
It results in the parity and cyclicity properties 
\be 
	B^\ssty{(N)}_n(\rho) = B^\ssty{(N)}_{n\pm N}(\rho)
	= (-1)^n B^\ssty{(N)}_{-n}(\rho) 
	= (-1)^n B^\ssty{(N)}_n(-\rho), \quad B^\ssty{(N)}_{n}(0)=\delta_{n,0},
		\lab{parities}
\ee	
which also hold for the continuous Bessel functions 
$J_n(\rho)$ of integer order \cite{Watson}.

A plane wave of wavenumber $\kappa$ along the $y$-axis 
in a Helmholtz medium that allows 
only $N$ equidistant directions of propagation on the circle,
can be obtained from \rf{Bdef-int} using the completeness 
relation \rf{ortho-complete} to expand the middle term
and write
\be   \begin{array}{rcl}
	\exp(\ii\rho\sin\varphi_m)= B^\ssty{(N)}_0(\rho)
		\!\!\!&+&\!\!\!\displaystyle 2\sum_{n=1}^{N-1} 
			B^\ssty{(N)}_{2n}(\rho)\cos(2n \varphi_m)\\
		{}\!\!\!&+&\!\!\! \displaystyle 2\ii\sum_{n=0}^{N-1} 
			B^\ssty{(N)}_{2n+1}(\rho)\sin((2n{+}1) \varphi_m),
			\end{array}	\lab{pl-save-exp}
\ee
showing how the discrete Bessel functions can take the
place of the continuous ones, {\it cf.}\ \cite[Eq.\ KU120(13)]{GR}. 	

In Ref.\ \cite{UW1} we proved analytically, and verified
numerically, that the following expressions for the discrete
Bessel functions are exact analogues of those valid for
continuous Bessel functions. Corresponding to 
\cite[WA44]{GR} for odd $N=:2j+1$, in Ref.\ \cite{UW1} we 
proved the linear relations involving the even and odd-$n$ 
discrete Bessel functions,
\bea 
	B_0(\rho) + \sum_{n=1}^j B_{2n}(\rho)\cos(2n\varphi_m) 
			&=& \cos(\rho\sin\varphi_m), \lab{lin1}\\
		\sum_{n=0}^j B_{2n+1}(\rho)\sin((2n{+}1)\varphi_m) 
			&=& \onehalf\sin(\rho\sin\varphi_m). \lab{lin2}
\eea
The quadratic formulas \cite[\S 7.6.2, Eq.\ (6)]{Graf-origin} 
associated to the name of Graf, were 
shown in Ref.\ \cite{Graf} to derive from the rotation of 
spherical harmonics through Wigner-$D$ functions, under 
contraction from the rotation to the Euclidean group.
These relations, of group-theoretical origin, retain their
validity under the discretization of the rotation subgroup, 
and lead to
\be 
	\sum_{n=-2j}^{2j} B_n(\rho)\,B_{n'-n}(\rho')
		= B_{n'}(\rho+\rho'),
			\lab{graff}
\ee
keeping in mind the parity property \rf{parities} 
for the negative $n$-indices in the sum
for odd $N$, addressing the vector rather than 
spin representations of the rotation group.

\begin{figure}[hbpt]
\centering  
\centerline{\includegraphics[scale=0.46]{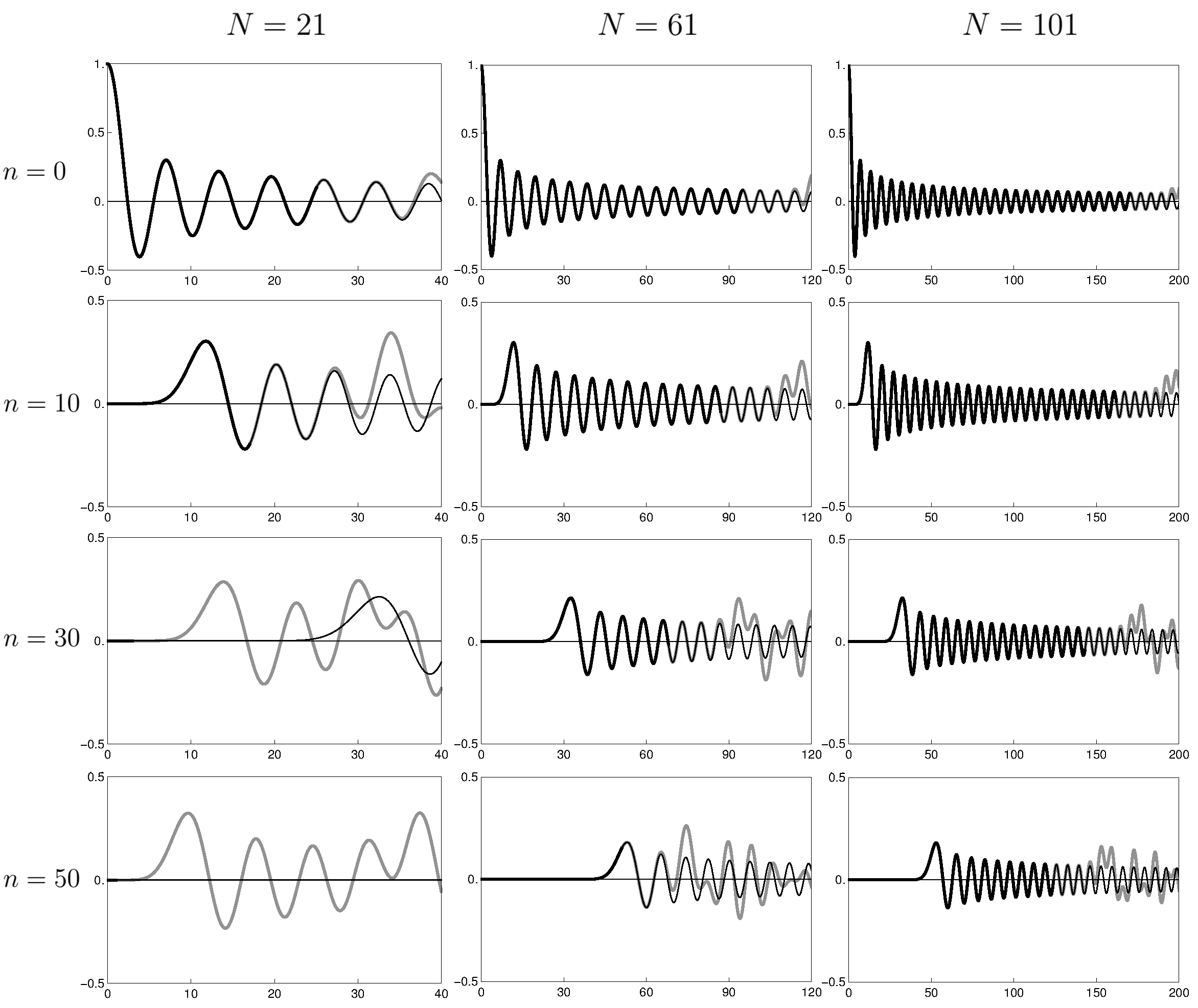}}
\caption[]{\footnotesize The `discrete' Bessel functions $B_{n}^\ssty{(N)}(\rho)$ on 
continuous intervals $0\le\rho\le(2N{-}1)$ (gray lines), {\it vs}.\ 
the `continuous' Bessel functions $J_n(\rho)$ (thin black lines), for orders 
$n\in\{0,\,10,\,30,\,50\}$ and point numbers $N\in\{21,\,61,\,101\}$.
Heavy black lines replace both where the `discrete' and 
the `continuous' Bessel functions differ by {\it less\/} 
than $10^{-16}$.} 
\label{fig:aproxms}
\end{figure}

In Fig.\ \ref{fig:aproxms} we essentially repeat
the figure in Ref.\ \cite{UW1} where we compared the 
discrete and continuous Bessel functions, $B^\ssty{(N)}_n(\rho)$
and $J_n(\rho)$, to support the claim that the approximation 
is indeed remarkable within an interval that is roughly 
$0\le n+\rho < N$. A similar set of figures is presented 
below for Mathiew functions.

Now, having $N$ basis functions 
$B^\ssty{(N)}_n(\rho)$, numbered by 
cyclic $n$ modulo $N$,  it is natural to inquire whether 
the argument $\rho$ can or should be also discretized 
to the $N$ integer values $\rho_k=k\in\{0,1,\ldots,N{-}1\}$. 
This was done in Ref.\ \cite{Biagetti-etal} while in \cite{UW1}
the plot in Fig.\ \ref{fig:aproxms} marked these points
and used them to define a kernel  $B^\ssty{(N)}_n(\rho_k)$ 
for a `discrete Bessel transform' between two $N$-vectors
of components $f_n$ and $\widetilde{f}_k$.
The fact is that while the angle $\varphi$ is discretized
naturally to $N$ points on the circle, the {\it radial\/} 
coordinate $\rho$ is {\it not\/} subject to a similarly
compelling set of points, but is valid and non-cyclic
over the complex $\rho$-plane.
The same discretization process for the angular ---but not
the radial--- coordinate will be applied to the Mathieu case below.


\section{Discrete Mathieu functions}
			\label{sec:three}

Elliptic coordinates on the plane generalize the previous
polar coordinates \rf{coord-rad}; they are defined in terms of
Cartesian coordinates through
\be 
	x=\cosh \varrho\,\cos\psi,\quad  y= \sinh \varrho\, \sin\psi,
		\qquad \varrho\in[0,\infty),\quad \psi\in(-\pi,\pi]={\cal S}^1.
		\lab{ell-coord}
\ee
where $(\varrho,\psi)$ are analogues of the previous polar 
coordinates $(r,\phi)$ for which we retain the names as `radial' 
and `angular' variables. For fixed $\varrho$ or for fixed $\psi$, 
the locus of points $(x,y)\in{\cal R}^2$ that satisfy
\be 
	{x^2}/{ \cosh^2\! \varrho} + {y^2}/{ \sinh^2\! \varrho}=1, \qquad
	{x^2}/{\cos^2\!\psi} - {y^2}/{ \sin^2\!\psi}=1,
		\lab{ell-hyper}
\ee
draw families of confocal ellipses or hyperbolas respectively.
At $\varrho=0$, $\psi\in{\cal S}^1$ draws twice the line between 
the two foci $(x,y)=(\pm 1,0)$ for $\psi=(0,\pi)$. 
The major and minor semi-axes of the ellipses are 
$\cosh \varrho$ and $\sinh\varrho$ respectively, so their
eccentricities are $1/\cosh \varrho$, that tend to
circles when $\varrho\to\infty$. On the other hand, for fixed
$\psi\in{\cal S}^1$ in each of the four quadrants, since
$\varrho\ge0$, only one of the four arms of the hyperbola is 
traversed. Thus we expect {\it four\/} parity cases out 
of the two reflections, across the $x$ and $y$ axes.
Compare this with the case of polar coordinates where
$r\ge0$ but all reflection axes are equivalent, so $(-1)^n$ 
in \rf{Bdef-int} provides the two Bessel parity cases.

\begin{figure}[hbpt]
\centering  
\centerline{\includegraphics[scale=0.2]{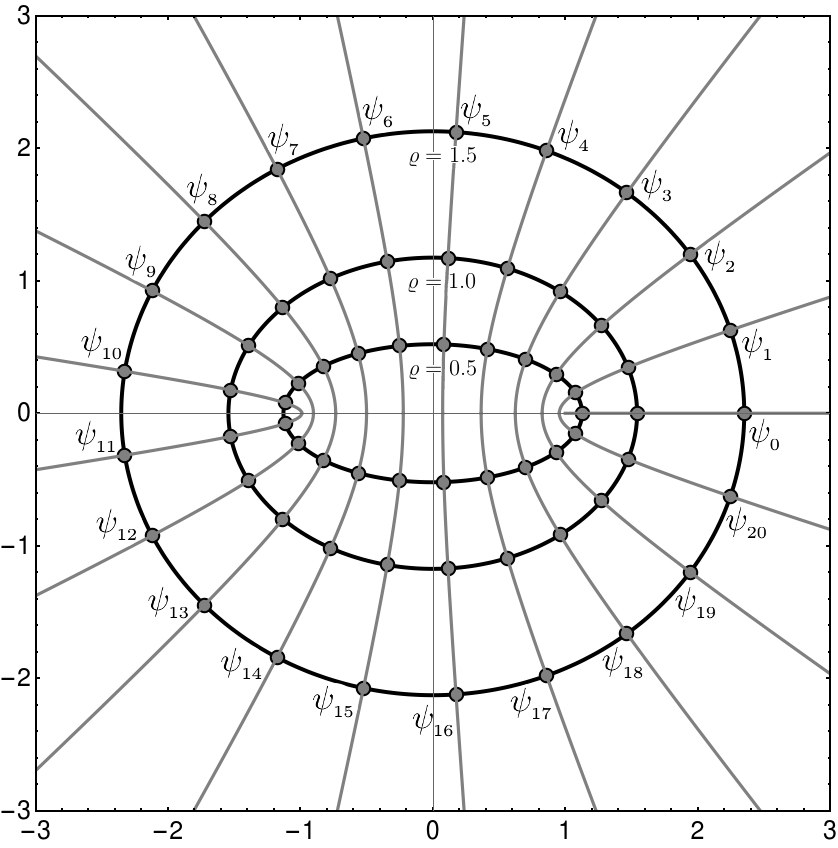}}
\caption[]{\footnotesize Set of equally-spaced discrete points on ellipses 
\rf{ell-coord} of the `angular' coordinates 
$\{\psi_m\}\in{\cal S}^1_\ssty{(N)}$ for $N=21$, 
and hyperbolas of the `radial' coordinate for
$\varrho\in\{0.5,\,1,\,1.5\}$.} 
\label{fig:elipses}
\end{figure}

The Helmholtz differential equation \rf{Helm-eq}, written in the
elliptic coordinates \rf{ell-coord}, is clearly separable,
\be 
	[(\partial^2_\varrho +\kappa^2 \cosh^2\!\varrho ) +
	 (\partial^2_\psi -\kappa^2\cos^2\!\psi)]\,f(\varrho,\psi)=0,
		\lab{Helm-ell}
\ee		
so that solutions can be written in the product form
$f(\varrho,\psi)\sim P(\varrho)\,\Psi(\psi)$. 
Dividing by $f$, one obtains two coupled equations 
in $\varrho$ and $\psi$, the latter is an eigenvalue
equation in the angular coordinate,
\be  
	(\partial^2_\psi  -2q\cos2\psi)\,\Psi(\psi,q)=\nu\,\Psi(\psi,q), 
		\qquad 	 q:=\tsty14\kappa^2,   \lab{Math-ang}
\ee
known as the Mathieu differential equation.  The angular 
coordinate $\psi$ is periodic and a well-known solution 
method consists in expanding solutions of \rf{Math-ang}
in the Fourier basis $\sim\exp(\ii n\psi)$ over all integer $n$.
This defines the Mathieu functions of the first kind 
${\rm ce}_{n}(\psi,q)$ and ${\rm se}_{n}(\psi,q)$ with
integer $n$ \cite{McLachlan}, characterized by a parity 
index $p\in\{0,1\}$ for even and odd cases \cite[Eqs.\ 8.61]{GR}, 
and distinct for even and odd indices. In a two-line expression
all cases can be written as
\be 
	\bigg[{{\rm ce}_{2n+p}(\psi,q)\atop{\rm se}_{2n+p}(\psi,q)}\bigg] =
			\sum_{s=0}^\infty\bigg[{A^{2n+p}_{2s+p}\cos((2s{+}p)\psi) 
			\atop B^{2n+p}_{2s+p}\sin((2s{+}p)\psi)}\bigg].
				\lab{class-Math}
\ee
The parities are even ${\rm ce}_n(-\psi,q)={\rm ce}_n(\psi,q)$,
odd ${\rm se}_n(-\psi,q)=-{\rm se}_n(\psi,q)$, and ${\rm se}_0(\psi,q)\equiv0$.
The coefficients $A^n_s,\,B^n_s$ are found introducing this expansion into
\rf{Math-ang} to find recursion relations \cite[Eqs.\ 8.62]{GR} that lead
to efficient numerical computation. For use below, we write
them using Fourier series as
\be 
	\bigg[{A^n_{s}\atop B^n_{s}}\bigg] =\frac1\pi
			\int_{{\cal S}^1} \dd\psi\,
			\bigg[{\cos(s\psi)\,{\rm ce}_n(\psi,q) 
			\atop \sin(s\psi)\,{\rm se}_n(\psi,q)}\bigg],
				\lab{AB-class-Math}
\ee
for $n\neq0$, while $A^n_0=(2\pi)^{-1}
\int_\ssty{{\cal S}^1}\dd\psi\,{\rm ce}_n(\psi,q)$, and
$B^n_0\equiv 0$. 
The Mathieu functions \rf{class-Math} are orthogonal under 
the inner product \rf{inn-prod-c} over the circle, 
$({\rm ce}_m,{\rm ce}_n)_\circ=\pi\delta_{m,n}$,
$({\rm se}_m,{\rm se}_n)_\circ=\pi\delta_{m,n}$
for $n\neq0$ ---zero otherwise, and
 $({\rm ce}_m,{\rm se}_n)_\circ=0$.

We now restrict the range of the angular coordinate 
$\psi$ from ${\cal S}^1$ to ${\cal S}^1_\ssty{(N)}$, shown
for the elliptic coordinates in Fig.\ \ref{fig:elipses},
in correspondence with the previous discrete phase functions 
in the Bessel case \rf{base-Phi}, and thus defining the `angular' 
{\it discrete Mathieu functions\/} of the
first type over $\psi_m\in{\cal S}^1_\ssty{(N)}$ as
\be 
	\bigg[{{\rm ce}^\ssty{(N)}_{2n+p}(\psi_m,q)
		\atop{\rm se}^\ssty{(N)}_{2n+p}(\psi_m,q)}\bigg]:=
			\sum_{s=0}^{N-1}\bigg[{a^{2n+p}_{2s+p}\cos((2s{+}p)\psi_m) 
				\atop b^{2n+p}_{2s+p}\sin((2s{+}p)\psi_m) }\bigg],
				\lab{disc-Math}
\ee
with coefficients $a^n_s,\,b^n_s$. The finite $N$-point 
Fourier transform approximates them through the 
replacement \rf{repl-lim} to 
the functions and coefficients $A^n_s,\,B^n_s$ 
of the continuous case in \rf{AB-class-Math}, as
\be 
	\bigg[{a^n_{s}\atop b^n_{s}}\bigg] :=\frac1N
			\sum_{m=0}^{N-1} 
			\bigg[{\cos(s\psi_m)\,{\rm ce}^\ssty{(N)}_n(\psi_m,q) 
			\atop \sin(s\psi_m)\,{\rm se}^\ssty{(N)}_n(\psi_m,q)}\bigg]
			\simeq\frac12 \bigg[{A^n_{s}\atop B^n_{s}}\bigg],
				\lab{ab-disc-Math}
\ee
for $s\neq 0$, while $a_0^n= A_0^n$, $b^n_0=0$, and also $b^0_s=0$. 

\begin{figure}[t]
\centering  
\centerline{\includegraphics[scale=0.276]{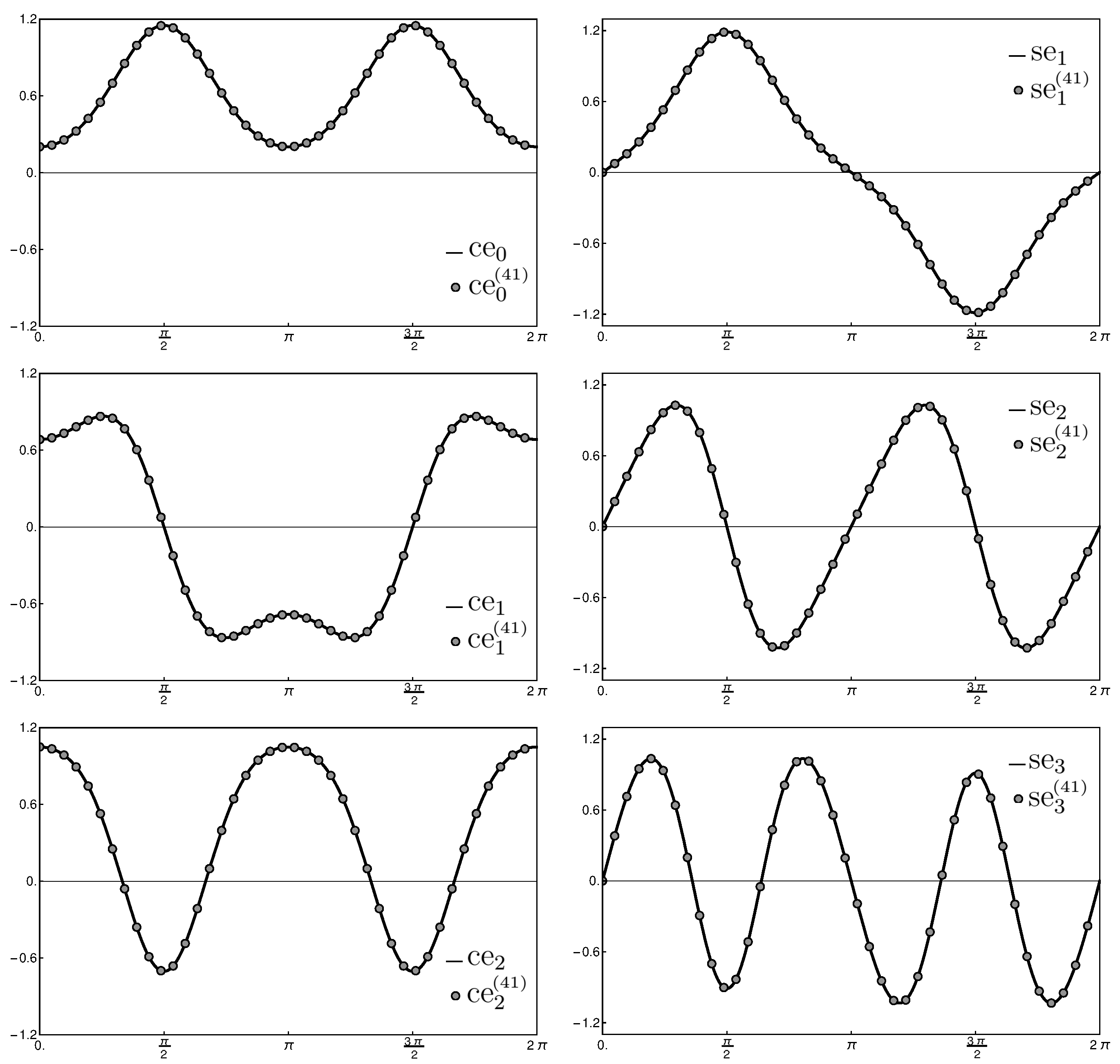}}
\caption[]{\footnotesize Discrete {\it vs.} continuous `angular' Mathieu 
functions for $N=41$, $q=2$. The values of the discrete functions
${\rm ce}^\ssty{(N)}_n(\psi_m,q)$ and ${\rm se}^\ssty{(N)}_n(\psi_m,q)$
at $\psi_m$, $0\le m\le N{-}1$, are indicated by circles. The continuous 
Mathieu functions ${\rm ce}_n(\psi,q)$ and ${\rm se}_n(\psi,q)$ 
are marked by black lines in their full range $0\le \psi<2\pi$.
Their difference is less than $10^{-16}$ for all points $\psi_m$.} 
\label{fig:Mathieu-angular}
\end{figure}

The last relation in \rf{ab-disc-Math} is an approximate equality, the
validity of which is contingent upon the numerical computation
and comparison between the lower- and upper-case coefficients within 
a range of their indices in, say, $0\le n,s\le N{-}1$, which is 
reflected in turn by the discrete and continuous Mathieu functions 
themselves. In Fig.\ \ref{fig:Mathieu-angular}
we compare a  sample of continuous angular Mathieu functions of the
first kind with their discrete approximations from Eq.\ \rf{disc-Math}. 
In favor of the thus defined discrete Mathieu functions, we note 
that they satisfy orthogonality relations under the discrete 
inner product \rf{disc-inn-prod}, namely
\be 
	({\rm ce}^\ssty{(N)}_n,{\rm ce}^\ssty{(N)}_{n'})_\ssty{(N)}=\onehalf N \delta_{n,{n'}},\!\quad
		({\rm se}^\ssty{(N)}_n,{\rm se}^\ssty{(N)}_{{n'}\neq0})_\ssty{(N)}
	=\onehalf N \delta_{n,{n'}},\!\quad ({\rm ce}^\ssty{(N)}_n,{\rm se}^\ssty{(N)}_{n'})_\ssty{(N)}=0.
		\lab{Math-orthog}
\ee	
By construction, the parities of the discrete Mathieu
functions are also even ${\rm ce}^\ssty{(N)}_n(-\psi_m,q)
={\rm ce}^\ssty{(N)}_n(\psi_m,q)$, or odd 
${\rm se}^\ssty{(N)}_n(-\psi_m,q)=-{\rm se}^\ssty{(N)}_n(\psi_m,q)$.

\begin{figure}[t]
\centering  
\centerline{\includegraphics[width=1.0\columnwidth]{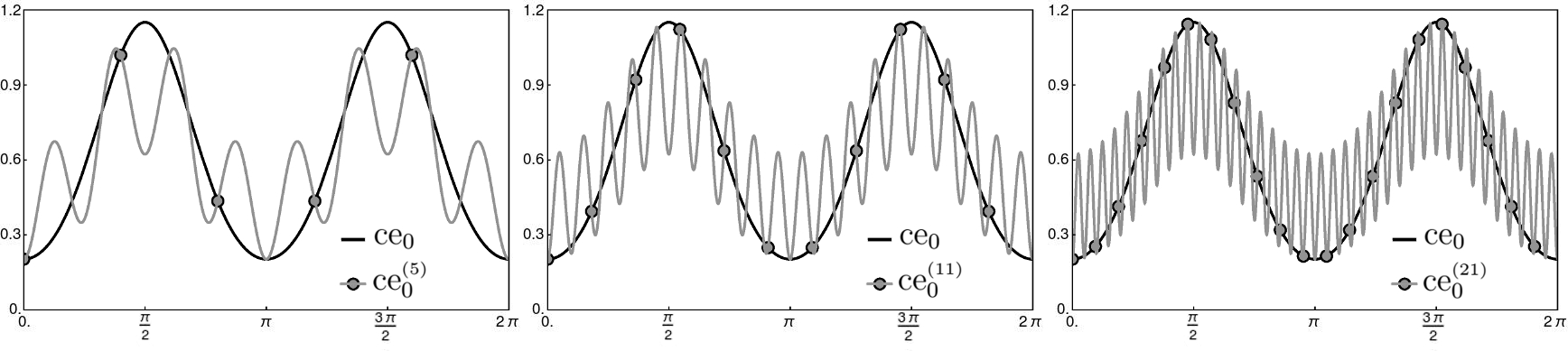}}
\caption[]{\footnotesize Comparison between the `discrete' Mathieu 
functions ${\rm ce}^\ssty{(N)}_0(\psi_m,q)$ whose arguments are 
continued to $\psi_m\in{\cal S}^1$ (gray line) {\it vs}.\ the 
`continuous' Mathieu function ${\rm ce}_n(\psi,q)$ (black line),
for point numbers $N\in\{5,11,21\}$ and $q=2$. The discrete points
$\psi_m\in{\cal S}^1_\ssty{(N)}$ lie at a subset of the 
intersections marked with circles.}
\label{fig:aproximaN}
\end{figure}

At this point it is illuminating to inquire into the manner in which
the discrete functions approximate the continuous ones. Consider for
example how ${\rm ce}^\ssty{(N)}_0(\psi_m,q)$, whose definition 
\rf{disc-Math} allows us to compute it for {\it continuous\/} 
$\psi_m\in{\cal S}^1$, matches ${\rm ce}_0(\psi,q)$ in the whole $\psi$ 
range. In Fig. \ref{fig:aproximaN} we do so for {\it small\/} $N$, noting
that where the continued $\psi_m$ lines of the former take their 
values, they intersect the properly continuous line of the latter;
although the two lines intersect also at other points, the two lines remain
notably distinct. As the figure shows, the approximation is {\it not\/}
valid over presumably small ranges {\it around\/} these intersections, 
but only {\it at\/} the prescribed $\psi_m=2\pi m/N$ points. We intend 
to elaborate on such and similar limits elsewhere.

Proceeding now as we did in \rf{disc-Helm}, but using the discrete Mathieu 
functions of the first kind ${\rm ce}^\ssty{(N)}_n(\psi_m,q)$ and 
${\rm se}^\ssty{(N)}_n(\psi_m,q)$ in place of the plain phase functions 
$\Phi^\ssty{(N)}_n(\phi_m)$, we again have Helmholtz solutions $f_n(\varrho,\psi_k)$ 
on the plane that are characterized by parities and sub-indices $n$, whose 
radial factor will be the discrete `radial' Mathieu functions of the 
{\it second\/} kind, to be indicated correspondingly as 
${\rm Ce}^\ssty{(N)}_n(\varrho,q)$ and ${\rm Se}^\ssty{(N)}_n(\varrho,q)$, 
\bea 
	{}\!\!\!\!\!\!\!\!\!\bigg[{f^\ssty{c}_{2n+p}(\varrho,\psi_k)\atop f^\ssty{s}_{2n+p+1}(\varrho,\psi_k)}\bigg] 
		&=& \frac1N \sum_{m=1}^{N-1} \bigg[{{\rm ce}^\ssty{(N)}_n(\psi_m,q)
			\atop {\rm se}^\ssty{(N)}_n(\psi_m,q)}\bigg]
				\exp[\ii\kappa(x\cos\psi_m+y\sin\psi_m)\lab{ff0}\\
		&=:& \bigg[{c_n(q)\, {\rm Ce}^\ssty{(N)}_n(\varrho,q)\,{\rm ce}^\ssty{(N)}_n(\psi_k,q)
		\atop s_n(q)\,{\rm Se}^\ssty{(N)}_n(\varrho,q)\,{\rm se}^\ssty{(N)}_n(\psi_k,q)}\bigg],
				\lab{ff-1}
\eea
where $c_n(q)$ and $s_n(q)$ are constants.
Using the elliptic coordinates with a discretized angular part,
$x(\varrho,\psi_k)=\cosh \varrho\*\cos \psi_k$ and 
$y(\varrho,\psi_k)=\sinh \varrho\*\sin \psi_k$ in \rf{ell-coord}, 
the phase exponent is then $\ii\kappa=2\ii\surd q$ times
$x\cos\psi_m+y\sin\psi_m = \cosh \varrho \*\cos \psi_k\*\cos\psi_m 
+ \sinh \varrho \*\sin \psi_k \*\sin\psi_m$.
As was done before in \rf{disc-Helm} and \rf{fase-B}, 
we extract the new discrete `radial' functions
using the orthogonality \rf{Math-orthog} of the 
previous discrete `angular' Mathieu functions, as
\be
	{}\!\!\!\!\!\begin{array}{r} \displaystyle 
	\bigg[{{\rm Ce}^\ssty{(N)}_{2n+p}(\varrho,q)\atop {\rm Se}^\ssty{(N)}_{2n+p+1}(\varrho,q)}	\bigg] 
		=\displaystyle \bigg[{1/N\,{c}_{2n+p}(q)\,{\rm ce}^\ssty{(N)}_{2n+p}(\psi_k,q)
		\atop 1/N\,{s}_{2n+p+1}(q)\,{\rm se}^\ssty{(N)}_{2n+p+1}(\psi_k,q)}\bigg]
			\sum_{m=0}^{N-1} 			
		\bigg[{{\rm ce}^\ssty{(N)}_{2n+p}(\psi_m,q) 
			\atop {\rm se}^\ssty{(N)}_{2n+p+1}(\psi_m,q)}\bigg]\\[10pt]
			{}\times \exp[2\ii\sqrt{q}\,(\cosh \varrho \*\cos \psi_k\*\cos\psi_m 
				+ \sinh \varrho \*\sin \psi_k \*\sin\psi_m)].	
				\end{array}   \lab{FF-disc-Math}
\ee
The coefficients in front of the summation will be now
determined through considering specific values for the `angular'
coordinate $\psi\leftrightarrow\psi_m$, comparing them with expressions
of the continuous Mathieu functions of the second kind obtained 
from integrals that are tabulated in Ref.\ \cite[\S 6.92]{GR}.
There, the exponential factors appear with only a single summand 
in the exponent, either sine or cosine. This occurs in 
\rf{FF-disc-Math} only for $\psi_m=0$ or $\onehalf\pi$, 
although the latter is not in the set ${\cal S}_\ssty{(N)}^1$ 
if $\psi_0=0$, since $N$ was assumed to be odd. 

\begin{figure}[hbpt]
\centering  
\centerline{\includegraphics[scale=0.12]{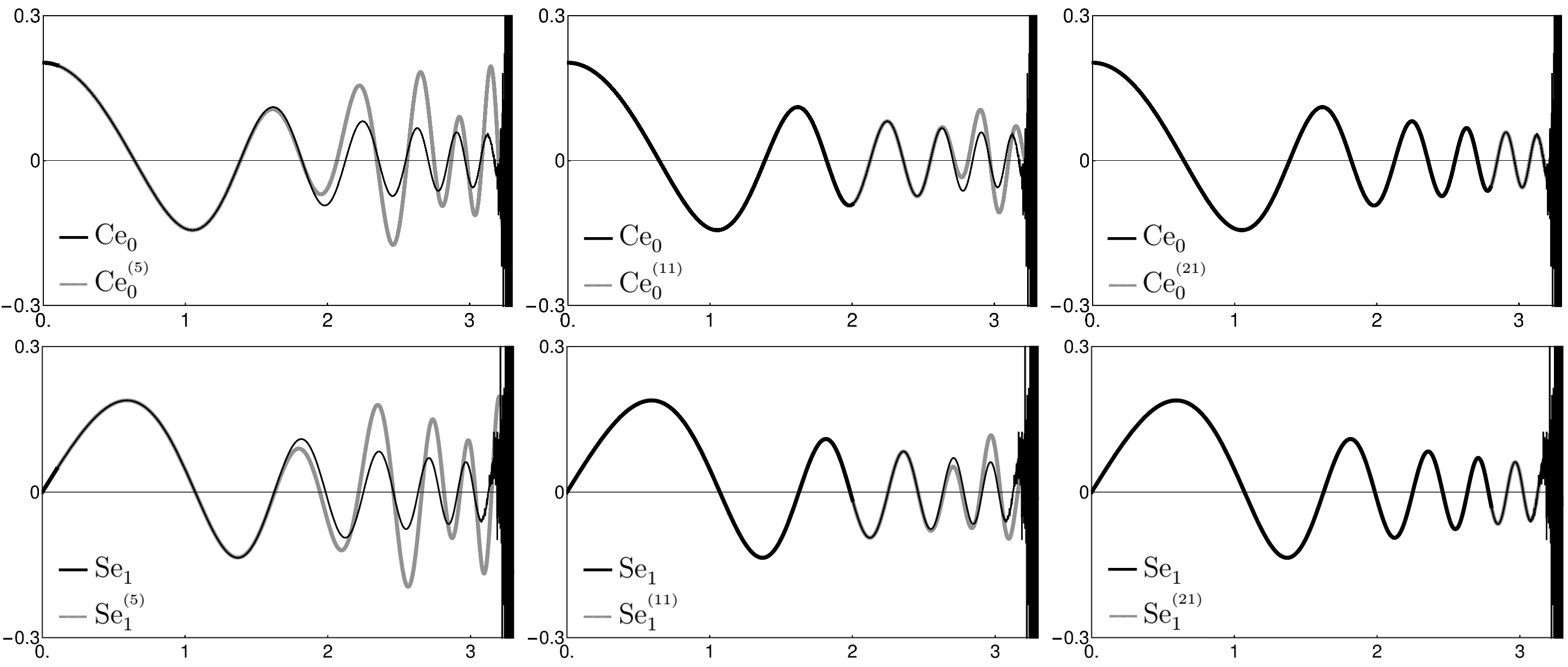}}
\caption[]{\footnotesize Discrete {\it vs.}\ continuous `radial' Mathieu functions in the 
interval $0\le\varrho<3.3$, for $N\in\{5,11,21\}$ and here for $q=2$. The `discrete' 
functions ${\rm Ce}^\ssty{(N)}_n(\varrho,q)$ and ${\rm Se}^\ssty{(N)}_n(\varrho,q)$
with the (continuous) argument $\varrho$ (gray line), is compared with the `continuous' 
functions ${\rm Ce}_n(\varrho,q)$ and ${\rm Se}_n(\varrho,q)$ (thin black line). As
before, where both coincide within $10^{-16}$ they are replaced by a thick black line.
The radial Mathieu functions, when computed with the commercial Mathematica algorithm, 
oscillate wildly after an upper value that decreases with increasing values of $q$.} 
\label{fig:Mathieu-radial}
\end{figure}

Let us first consider the case of even parity $p=0$ and the angle 
$\psi=\onehalf\pi$ in \rf{FF-disc-Math}, where the previous remark 
applies. Based on the close approximation between the discrete 
and continuous Mathieu functions, we may simply replace the latter
for the former, so that the two lines in that expression read
\be 
	\bigg[{{\rm Ce}^\ssty{(N)}_{2n}(\varrho,q)
		\atop {\rm Se}^\ssty{(N)}_{2n+1}(\varrho,q)}\bigg] 
	=\bigg[{K^\ssty{c}_{2n} \atop K^\ssty{s}_{2n+1}}\bigg]
		\sum_{m=0}^{N-1} 			
		\bigg[{{\rm ce}^\ssty{(N)}_{2n}(\psi_m,q) 
			\atop {\rm se}^\ssty{(N)}_{2n+1}(\psi_m,q)}\bigg]
			 \exp(2\ii\sqrt{q}\,\sinh\varrho \sin\psi_m). 
		\lab{DiscCe0}
\ee 	
When this summation formula is compared
with the integral expressions tabulated 
in \cite[\S 6.92]{GR}, namely
\be 
	\begin{array}{r} \displaystyle 
	\bigg[{{\rm Ce}_{2n}(\varrho,q)\atop {\rm Se}_{2n+1}(\varrho,q)}\bigg] 
		=\bigg[{{\rm ce}_{2n}(0,q)/2\pi\,A_0^{2n}  
				\atop -\ii{\rm se}'_{2n+1}(0,q)/2\pi\,B_1^{2n+1}\!\!\sqrt q}\bigg]
				\qquad\qquad\qquad\qquad{}\\[10pt]
				\displaystyle{}\times  \int_{{\cal S}^1} \dd\psi\,		
		\bigg[{{\rm ce}_{2n}(\psi,q) \atop {\rm se}_{2n+1}(\psi,q)}\bigg]
			\exp(2\ii\sqrt{q}\,\sinh \varrho \sin\psi),\end{array}
			\lab{ContCe0}
\ee
we conclude that the constants in the summation \rf{DiscCe0},
after identifying $2\pi\leftrightarrow N$, $A_0^{2n}=a_0^{2n}$
and $B_1^{2n+1}\simeq 2b_1^{2n+1}$, are
\be 
	K^\ssty{c}_{2n} = \frac{{\rm ce}_{2n}(0,q)}{a_0^{2n} N},\qquad
	K^\ssty{s}_{2n+1}=\frac{-\ii\,{\rm se}'_{2n+1}(0,q)}{2 b_1^{2n+1} N\surd q}.
		\lab{consts}
\ee
where ${\rm se}'_n(0,q):= \dd \,{\rm se}_n(\psi,q)/\dd\psi |_{\psi=0}$.
In Fig.\ \ref{fig:Mathieu-radial} we compare a sample of the discrete and 
continuous `radial' Mathieu functions, noting that the two lines are
quite coincident in the range $\varrho\in[0,\pi)$, but that the discrete 
approximant oscillates wildly beyond $\pi$. Again, here we can only 
justify this statement numerically.

Next we consider the case of odd parity $p=1$ at the 
value $\psi=\psi_0=0$. The upper line  in \rf{DiscCe0} reads
\be 
	{\rm Ce}^\ssty{(N)}_{2n+1}(\varrho,q) = K^\ssty{c}_{2n+1}
		\sum_{m=0}^{N-1}{\rm ce}^\ssty{(N)}_{2n+1}(\psi_m,q)
			\exp(2\ii\sqrt q\cosh\varrho\cos\psi_m),
			\lab{CeN2n1}
\ee
that we compare with the integral for the continuous Mathieu 
functions of the second kind in \cite[\S 6.92]{GR}, namely
\be 
	{\rm Ce}_{2n+1}(\varrho,q)
		= \frac{\ii\,{\rm ce}'_{2n+1}(\onehalf\pi,q)
			}{2\pi A_1^{2n+1} \sqrt q} \int_{{\cal S}^1}\dd\psi\,
		{\rm ce}_{2n+1}(\psi,q)\exp(2\ii\sqrt q\cosh\varrho\cos\psi),
			\lab{Ce2n1}
\ee
where ${\rm ce}'_{2n+1}(\psi,q)$ is the derivative of
the Mathieu function.  Again exploiting the correspondences 
\rf{repl-lim}, $2\pi\leftrightarrow N$ 
and $A_1^{2n+1}\simeq 2a_1^{2n+1}$, we conclude the constant 
in \rf{CeN2n1} to be
\be 
	K^\ssty{c}_{2n+1} = \frac{\ii\,{\rm ce}'_{2n+1}(\onehalf\pi,q)
		}{2a_1^{2n+1} N \surd q}.
			\lab{c2n1}
\ee

The remaining case to be considered is that of odd parity and
even index, namely for ${\rm Se}^\ssty{(N)}_{2n+2}(\varrho,q)$.
This presents a problem though, because the summation \rf{FF-disc-Math}
is identically zero for both $\psi_k=0$ and $\onehalf\pi$ due to
the parities of the terms in the sum.
It is different from zero for $0<\psi_k<\onehalf\pi$ however,
so if we choose $\psi_k=\frac14 \pi$, where both summands in the
exponent appear as $1/\surd2$, we can write
\be 
	\begin{array}{r} \displaystyle
	{\rm Se}^\ssty{(N)}_{2n+2}(\varrho,q) = K^\ssty{s}_{2n+2}
		\sum_{m=0}^{N-1}{\rm se}^\ssty{(N)}_{2n+2}(\psi_m,q)
			\qquad\qquad\qquad\qquad{}\\[5pt] 	\displaystyle{}\times 
		\exp[\ii\sqrt{2q}\,(\cosh\varrho\cos\psi_m+\sinh\varrho\sin\psi_m)].
			\end{array}  \lab{CeN2n1}
\ee
For the corresponding continuous case, we could not find a 
corresponding integral in \cite[\S 6.92]{GR}, so we cannot give
a closed expression for the coefficient $K^\ssty{s}_{2n+2}$ 
in \rf{CeN2n1}. The lack of a similar plane-wave integral
expression for the continuous Mathieu functions 
${\rm Se}_{2n+2}(\varrho,q)$ has been noted also in 
Ref.\ \cite{Mathieu-LeyKoo} without explanation.
However, we have checked numerically that the simile 
of the discrete to continuous functions approximation
provided by
\be 
		{\rm Se}^\ssty{(N)}_{2n+2}(\varrho,q) 
		 \simeq 
			-\ii\,{\rm se}_{2n+2}(\ii\varrho,q) 
			= {\rm Se}_{2n+2}(\varrho,q),
				\lab{Seise}
\ee
which is an equality for continuous functions, {\it cf}.\ 
\cite[8.611.4, 8.631.4]{GR}. For $0<\varrho<2$
the difference in \rf{Seise} less than $10^{-14}$. We
should note that generally the discrete `radial' and `angular'
Mathieu functions for pure imaginary arguments are {\it not\/} 
related to similar equalities of their continuous integral 
expressions, because the summation definitions in \rf{disc-Math} 
involve hyperbolic functions. In particular, say,
\be 
	{\rm Se}^\ssty{(N)}_{2n+p+1}(\varrho,q) \neq
	-\ii{\rm se}^\ssty{(N)}_{2n+p+1}(\ii\varrho,q)
	= \sum_{m=0}^{N-1} b^{2n+p+1}_{2m+p+1}
		\sinh[(2m{+}p{+}1)\varrho]. 
			\lab{noes}
\ee


\section{Concluding remarks}   \label{sec:four}

The expansion of Helmholtz plane waves in 
series of radial Bessel and angular trigonometric 
functions has its discrete analogue in Eq.\ 
\rf{pl-save-exp}, which tells us that the wavefield
due to a finite number  $N$ of plane waves at
equidistant direction angles can be expanded
in discrete Bessel radial functions and 
corresponding trigonometric angular functions.
A similar statement will hold when the wavefield
is expanded in discrete Mathieu functions with
the phases determined by the points on an ellipse
as depicted in Fig.\ \ref{fig:elipses}. 
Conceivably such fields can be produced in 
resonant two-dimensional micro-cavities fed by
a number of activation channels.

We recognize that the full treatment and 
exploration of properties for the discrete
Bessel and Mathieu function presented here
is not exhaustive, but that it should be sufficient
to indicate that the approximation method 
consisting in the replacement of a continuous 
closed subgroup of the symmetry group of a 
partial differential equation by a finite 
discrete group is definitely of interest. 
In the present case of two
dimensions, the orthogonal group was reduced 
to the dihedral group. In three dimensions, 
the symmetry Euclidean symmetry group could 
reduce its three-dimensional rotation subgroup
by any of its polyhedral subgroups, whose 
functions may serve to describe wavefields
with a corresponding subset of 
wave propagation directions. Here we have presented
a set of exact relations, others whose 
approximation closeness was estimated through
numerical computation, and others that have
been only suggested by that approach, and for
which we expect to present further results from
ongoing work.


\section*{Acknowledgments}

We thank the support of the Universidad Nacional Aut\'onoma
de M\'exico through the PAPIIT-DGAPA project AG100120
{\it \'Optica Matem\'atica}.

\end{document}